\documentstyle[baasar]{article}

\begin{document}

\title{The Rise and Citation Impact of astro-ph in Major Journals}

\author{Travis~S.~Metcalfe}

\affil{High Altitude Observatory, National Center for Atmospheric Research,
       P.O.~Box 3000, Boulder, CO 80307-3000 USA}

\begin{abstract}

The rise in the use of the arXiv preprint server (astro-ph) over the past 
decade has led to a major shift in the way astronomical research is 
disseminated. Schwarz \& Kennicutt (2004) recently found that {\it 
Astrophysical Journal} papers posted to astro-ph are cited roughly twice 
as often as papers that are not posted, suggesting that the preprint 
server has become the primary resource for many astronomers to keep up 
with the literature. I describe a simple method to determine the adoption 
rate and citation impact of astro-ph over time for any journal using 
NASA's Astrophysics Data System (ADS). I use the ADS to document the rise 
in the adoption of astro-ph for three major astronomy journals, and to 
conduct a broad survey of the citation impact of astro-ph in 13 different 
journals. I find that the factor of two boost in citations for astro-ph 
papers is a common feature across most of the major astronomy journals.

\end{abstract}

%\keywords{sociology of astronomy -- astronomical data bases: miscellaneous}}

\section{Motivation}

Schwarz \& Kennicutt (2004) recently conducted a detailed study of papers 
published in the {\it Astrophysical Journal} (ApJ), designed in part to 
document the adoption rate and citation impact of the arXiv preprint 
server (astro-ph; Ginsparg 2001). The database they constructed for this 
study is a rich source of demographic and citation information for the 
ApJ, but the two most interesting results were specific to astro-ph. They 
found that the adoption rate of astro-ph varied widely between subfields 
of astronomy, with nearly universal adoption (95\%) in cosmology and the 
lowest adoption (22\%) in solar astrophysics and planetary science by 
2002. This difference among subfields is significant because they also 
found that papers posted to astro-ph are cited about twice as often as 
papers that are not posted. This boost in citation rates was more than a 
factor of 10 for cosmology papers, and still a factor of 1.8 for papers in 
solar astrophysics and planetary science.

Considering the main findings of Schwarz \& Kennicutt about ApJ papers, I 
wondered whether the other major journals would show similar trends. Is 
astro-ph as widely used by author communities in other parts of the world? 
Does astro-ph produce a comparable impact on the citation rates for papers 
published in the other journals? Answering these narrow questions for a 
large sample of journals is considerably easier than the deep study 
undertaken by Schwarz \& Kennicutt for the ApJ. The Astrophysics Data 
System (ADS; Kurtz et al.~2000) recently added links to the arXiv preprint 
records of journal papers that have also been posted to astro-ph. This new 
feature, combined with the citation data and search filtering capability 
of ADS, allows users to measure the adoption rate and citation impact of 
astro-ph for each journal with just a few clicks.

In this short paper, I document the rise in the adoption rate of astro-ph 
for three major astronomy journals published in North America (ApJ), the 
UK (MNRAS), and continental Europe (A\&A). I also determine the citation 
impact of astro-ph on papers published in 2002 for 13 journals spanning 
the globe. I describe in section~\ref{sec2} how to generate these 
statistics for any journal using the ADS, and I validate the method by 
comparing it to the Schwarz \& Kennicutt sample. I present and discuss the 
main results in section~\ref{sec3}, and I summarize the major conclusions 
in section~\ref{sec4}.

\section{Methodology\label{sec2}}

For any given range of dates and for any particular journal in the ADS, 
statistics on the fraction of papers posted to astro-ph (the adoption 
rate) and the resulting impact on the citation rate (the citation impact) 
can be generated with two simple queries. For example, generating 
statistics that can be compared to the Schwarz \& Kennicutt sample 
requires the following: (1) Enter a {\tt Publication Date} range between 
{\tt 7/1999} and {\tt 12/1999}. (2) Under the filters section {\tt Select 
References From}, choose {\tt All refereed publications} and type the 
journal code ``{\tt ApJ..}'' in the {\tt select/deselect publications} 
field. (3) For the first query, set {\tt Select References With} to {\tt A 
bibliographic entry} (the default). For the second query, set it to {\tt 
All of the following} and check {\tt ArXiv Preprint}. (4) Under the 
sorting section, select {\tt Sort by citation count}. In the query 
results, just above the bibliographic entries, the first query yields the 
total number of ApJ papers published in this date range and the total 
number of citations to those papers in the ADS database (which is not 
complete, but includes data from all of the major journals beginning in 
1999). The second query yields the corresponding numbers for the subset of 
these papers that have also been linked to an astro-ph record.

% FIGURE 1 %%%%%%%%%%%%%%%%%%%%%%%%%%%%%%%%%%%%%%%%%%%%%%%%%%%%%%%%%%%%%%
\begin{figure*}
\plotfiddle{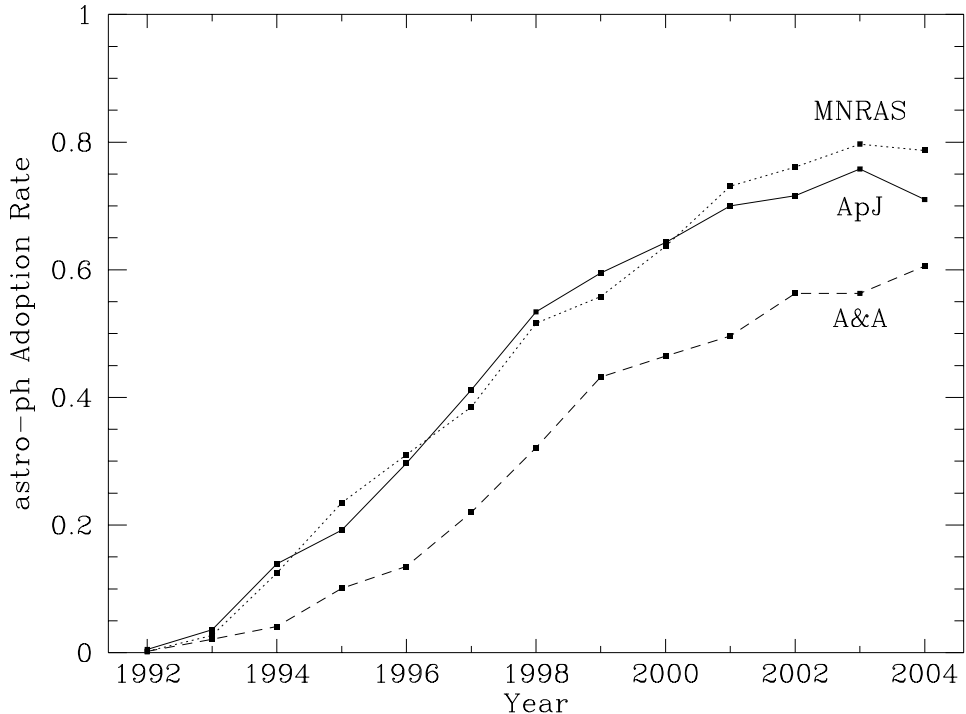}{3.5in}{270}{50}{50}{-210}{285}
\caption{The rise of astro-ph in three major journals published in the UK 
(MNRAS), North America (ApJ), and continental Europe (A\&A).\label{fig1}}
\end{figure*}
%%%%%%%%%%%%%%%%%%%%%%%%%%%%%%%%%%%%%%%%%%%%%%%%%%%%%%%%%%%%%%%%%%%%%%%%%

In this example, the first query results in {\tt Total number selected: 
1182, Total citations: 29904}, while the second query gives {\tt Total 
number selected: 706, Total citations: 22623}. This means that the totals 
for non-astro-ph papers are 476 and 7281 respectively. Since these queries 
return papers published as Letters and Main Journal articles together, and 
because additional citations have now been added to ADS, the results are 
not strictly comparable to the Schwarz \& Kennicutt sample. Even so, the 
fraction of ApJ articles posted to astro-ph from these queries 
($706/1182=60\%$) is nearly identical to that found for Main Journal 
articles by themselves (61\%; Schwarz \& Kennicutt 2004, their Table 4). 
In addition, the mean number of citations for astro-ph papers 
($22623/706=32.0$) compared to non-astro-ph papers ($7281/476=15.3$) from 
these queries is larger by a factor of 2.09, which is similar to the 
overall factor of 2.05 found by Schwarz \& Kennicutt (2004, their Table 
6). This broad agreement suggests that the ADS can by itself generate 
useful statistics on the adoption rate and citation impact of astro-ph.

For papers published in 2002, I repeated the procedure described above for 
13 different journals spanning the globe and covering a broad range of ISI 
impact factors\footnote{The Institute for Scientific Information (ISI) 
impact factor, tabulated annually for each journal, measures the average 
number of citations per paper per year during the first two years after 
publication.}. For three major astronomy journals (ApJ, MNRAS, and A\&A) I 
also tracked the annual rise in the adoption rate of astro-ph from 1992 to 
2004, to look for differences in the preprint posting habits of these 
author communities.

\section{Results \& Discussion\label{sec3}}

The rise in the use of astro-ph over the past decade has led to a major 
shift in the way astronomical research is disseminated, and it has had a 
dramatic impact on the citation rates of posted papers as a consequence. 
In Figure~\ref{fig1}, I show the fraction of papers posted to astro-ph for 
three major astronomy journals from 1992 (astro-ph began in April of that 
year) to 2004. For all three journals, the adoption rate of astro-ph grew 
steadily into the late 1990's and then started to level off over the past 
few years. Authors of papers appearing in the ApJ and MNRAS have adopted 
astro-ph at roughly the same rate over this time period, while authors 
publishing in A\&A have generally been slower to adopt it. Current 
adoption rates for MNRAS and the ApJ have leveled off near 70-80\%. As 
noted by Schwarz \& Kennicutt, this is similar to the fraction of authors 
who allow their articles to be posted in preprint form on the ApJ website 
(without any additional work by the author). This service was initiated by 
the ApJ in 2002, and may have actually contributed to the limited growth 
in astro-ph submissions in recent years by authors who considered it 
equivalent to astro-ph (MNRAS did not start a similar service until 2004). 
The adoption rate of astro-ph among A\&A authors has grown more slowly and 
currently stands near 60\%, a value the ApJ and MNRAS both reached in 
2000. Within the next 5 years, the use of astro-ph seems likely to become 
almost universal among authors of papers in these major astronomy 
journals.

% TABLE 1 %%%%%%%%%%%%%%%%%%%%%%%%%%%%%%%%%%%%%%%%%%%%%%%%%%%%%%%%%%%%%%%
\begin{table*}
\centering
\caption{The citation impact of astro-ph for various journals.\label{tab1}}
\begin{tabular}{lcrcrccrccccc}
\tableline
Journal~~~~~~~~~~&&Impact&&\multicolumn{2}{c}{astro-ph}&&\multicolumn{2}{c}{non-astro-ph}&&Adoption&&Citation\\
&&Factor&&\#&$\langle$citations$\rangle$&&\#&$\langle$citations$\rangle$&&Rate&&Impact\\
\tableline
  Nature\dotfill && 30.432 &&   13 & $33.4\pm9.4$ &&   97 &  $6.2\pm0.7$ && 0.12 && $5.4\pm1.6$ \\ 
 Science\dotfill && 26.682 &&   13 & $31.9\pm9.0$ &&   31 &  $6.4\pm1.2$ && 0.30 && $5.0\pm1.7$ \\ 
     ApJ\dotfill &&  6.187 && 1670 & $14.9\pm0.4$ &&  661 &  $7.1\pm0.3$ && 0.72 && $2.1\pm0.1$ \\ 
      AJ\dotfill &&  5.119 &&  327 & $13.9\pm0.8$ &&  182 &  $7.8\pm0.6$ && 0.64 && $1.8\pm0.2$ \\ 
    ApJS\dotfill &&  4.749 &&   68 & $18.5\pm2.3$ &&   58 &  $7.3\pm1.0$ && 0.54 && $2.5\pm0.5$ \\ 
   MNRAS\dotfill &&  4.671 &&  809 & $13.1\pm0.5$ &&  254 &  $3.8\pm0.3$ && 0.76 && $3.5\pm0.3$ \\ 
    A\&A\dotfill &&  3.781 && 1034 & $ 9.4\pm0.3$ &&  802 &  $4.6\pm0.2$ && 0.56 && $2.0\pm0.1$ \\ 
  Icarus\dotfill &&  3.009 &&   11 &  $4.1\pm1.4$ &&  235 &  $1.9\pm0.2$ && 0.04 && $2.1\pm0.7$ \\ 
    PASP\dotfill &&  2.830 &&   62 & $10.9\pm1.4$ &&   76 &  $3.4\pm0.4$ && 0.45 && $3.2\pm0.6$ \\ 
    PASJ\dotfill &&  1.996 &&   67 & $ 5.9\pm0.8$ &&   63 &  $3.8\pm0.5$ && 0.52 && $1.6\pm0.3$ \\ 
    SoPh\dotfill &&  1.875 &&   11 & $ 2.8\pm1.0$ &&  168 &  $3.6\pm0.3$ && 0.06 && $0.8\pm0.3$ \\ 
    PASA\dotfill &&  0.898 &&   24 & $ 3.8\pm0.9$ &&   41 &  $2.0\pm0.4$ && 0.37 && $1.9\pm0.6$ \\ 
  Ap\&SS\dotfill &&  0.383 &&   47 & $ 1.6\pm0.3$ &&  265 &  $0.5\pm0.1$ && 0.15 && $2.9\pm0.6$ \\ 
\tableline
\end{tabular}
\end{table*}
%%%%%%%%%%%%%%%%%%%%%%%%%%%%%%%%%%%%%%%%%%%%%%%%%%%%%%%%%%%%%%%%%%%%%%%%%

A broad survey of the adoption rate and citation impact of astro-ph on 
papers published in 2002 for a wide variety of journals is summarized in 
Table~\ref{tab1}. The survey includes high-impact multidisciplinary 
journals like {\it Nature} and {\it Science}, as well as a selection of 
astronomy journals published in various parts of the world with a wide 
range of impact factors. The topical journals {\it Solar Physics} and {\it 
Icarus} were included to try to elucidate the low adoption rate found by 
Schwarz \& Kennicutt for ApJ papers in the subfield that included solar 
astrophysics and planetary science. The most striking feature of 
Table~\ref{tab1} is that in almost every journal surveyed, papers posted 
to astro-ph are cited significantly more often than papers that are not 
posted. For most of the major astronomy journals, astro-ph papers are 
cited between 1.6 and 3.5 times as often as non-astro-ph papers. The 
median boost is about a factor of 2, which is comparable to what was found 
by Schwarz \& Kennicutt for ApJ papers.

Astronomy papers published in the high-impact journals get an even larger 
boost from being posted to astro-ph. Despite a significantly lower 
adoption rate, astronomy papers appearing in {\it Nature} and {\it 
Science} that are also posted to astro-ph are cited about 5 times more 
often than papers that are not posted. This extra boost may come from a 
kind of ``brand recognition'' associated with these high-impact journals, 
capturing an even broader audience than an astro-ph paper published in one 
of the other journals. High-impact journal papers that are not posted to 
astro-ph have citation rates comparable to non-posted ApJ papers, and are 
actually cited less often than major astronomy journal papers that are 
posted to astro-ph. Similarly, astro-ph papers that appear in journals 
with the lowest impact factors still get a citation boost over 
non-astro-ph papers near the average. As expected, the {\it absolute} 
citation rate is roughly correlated with the ISI impact factor, but the 
boost in citations due to astro-ph does not change by much across a wide 
selection of astronomy journals.

The only papers in this survey that did not receive a significant boost 
from being posted to astro-ph were those published in {\it Solar Physics} 
(SoPh). Schwarz \& Kennicutt noted that the ApJ papers with the lowest 
adoption rate of astro-ph were in the solar system (SS) subfield, which 
included solar astrophysics and planetary science. The low adoption rate 
by itself may not explain the lack of a citation impact for SoPh since the 
adoption rate is even lower for {\it Icarus} papers, which are still cited 
twice as often when posted to astro-ph. An alternative explanation is that 
relatively few solar astrophysicists learn about new research through 
astro-ph. It is simply not yet part of the culture in this subfield.

\section{Conclusions\label{sec4}}

Across a wide selection of major astronomy journals, papers that are 
posted to astro-ph are cited about twice as often as papers that are not 
posted. As the single source containing most of the new research to be 
published in refereed journals around the world, the astro-ph preprint 
server appears to be the method that most astronomers now use to keep up 
with the literature. If citation rates are any indication of the 
assimilation of new research by the astronomical community, then astro-ph 
seems to be the best single form of advertising available. Editors who 
want to maximize the impact factor of their journals should encourage 
authors to post their preprints to astro-ph, and authors in subfields 
where astro-ph has not yet been adopted should consider the advantages 
that other subfields have already discovered.

\acknowledgements
This research has made use of NASA's Astrophysics Data System and was 
supported by the National Science Foundation through an Astronomy \& 
Astrophysics Postdoctoral Fellowship under award AST-0401441.

\end{document}